%% file: main.tex
\documentclass[manuscript,nonacm]{acmart}

\AtBeginDocument{%
  }

\input{my_style}

\input{my_notation}

\begin{document}

\title{From Transparency to Accountability and Back: A Discussion of Access and Evidence in AI Auditing}

 \author{Sarah H. Cen}

 \email{shcen@stanford.edu}
 \affiliation{%
	\institution{Stanford University}
	\city{Palo Alto}
	\state{California}
	\country{USA}
}

 \author{Rohan Alur}

 \email{ralur@mit.edu}
 \affiliation{%
   \institution{Massachusetts Institute of Technology}
   \city{Cambridge}
   \state{Massachusetts}
   \country{USA}
 }

\begin{abstract}

\input{sections/abstract}
\end{abstract}

\maketitle

\input{sections/intro}

\input{sections/related_work}

\input{sections/aedt_case_study}

\input{sections/black_box_2}

\input{sections/framework}

\input{sections/challenges}

\begin{acks}
	Thank you to Manish Raghavan, Aleksander Madry, Martha Minow, Cosimo Fabrizio, and James Siderius for their valuable feedback on this work. 
 	The authors gratefully acknowledge funding from the MIT-IBM project on Causal Representation, a Stephen A. Schwarzman College of Computing Seed Grant, the
 	National Science Foundation (NSF) grant CNS-1955997,
 	and the
 	Air Force Research Laboratory (AFOSR) grant FA9550-23-1-0301.
\end{acks}

\bibliographystyle{ACM-Reference-Format}
\bibliography{references}

\appendix

\input{sections/tech_related_work}

\input{sections/additional_related_work}

\end{document}

%% file: my_style.tex
\usepackage{algorithm} 
\usepackage{algpseudocode}

\usepackage[]{color-edits}

\addauthor{shc}{red}
\addauthor{ra}{blue}
\usepackage{todonotes}

\usepackage{setspace}
\usepackage{enumitem}

\usepackage{hyperref}
\usepackage{cleveref}

\usepackage{amsthm}
\newtheorem{example}{Example}

\newtheorem{definition}{Definition}

\usepackage{graphicx}
\usepackage{caption} %
\usepackage{subcaption}
\usepackage{accents} %

\usepackage{booktabs}
\usepackage{pdflscape}
\usepackage{longtable}
\usepackage{array}
\newcolumntype{L}[1]{>{\raggedright\arraybackslash}p{#1}}
\usepackage{footnote}
\makesavenoteenv{tabular}
\usepackage{threeparttable-custom}

%% file: my_notation.tex
\newcommand{\cX}{\mathcal{X}}
\newcommand{\cY}{\mathcal{Y}}
\newcommand{\bbE}{\mathbb{E}}

\newcommand{\cF}{\mathcal{F}}

\newcommand{\reals}{\mathbb{R}}

%% file: sections/abstract.tex
Artificial intelligence (AI) is increasingly intervening in our lives, raising widespread concern about its unintended and undeclared side effects. These developments have brought attention to the problem of \emph{AI auditing}: the systematic evaluation and analysis of an AI system, its development, and its behavior relative to a set of predetermined criteria. 
Auditing can take many forms, including pre-deployment risk assessments, ongoing monitoring, and compliance testing. It plays a critical role in providing assurances to various AI stakeholders, from developers to end users. Audits may, for instance, be used to verify that an algorithm complies with the law, is consistent with industry standards, and meets the developer's claimed specifications. 
However, there are many operational challenges to AI auditing that complicate its implementation. 

In this work,  we examine a key operational issue in AI auditing: 
what type of access to an AI system is needed to perform a meaningful audit?
Addressing this question has direct policy relevance, as it can inform AI audit guidelines and requirements.  
We begin by discussing the factors that auditors balance when determining the appropriate type of access, 
and unpack the benefits and drawbacks of four types of access. 
We conclude that, at \emph{minimum}, {black-box access}---providing query access to a model without exposing its internal implementation---should be granted to auditors. In particular, we argue that black-box access effectively balances concerns related to proprietary technology, data privacy, audit standardization, and audit efficiency.
We then suggest a framework for determining how much \emph{further} access (in addition to black-box access) to provide to auditors. 
We show that auditing can be cast as a natural hypothesis test and argue that this framing provides clear and interpretable guidance on the implementation of AI audits. 
In particular, we draw parallels between aspects of hypothesis testing and those of legal procedure, such as legal presumption and burden of proof. 
As a result, hypothesis testing provides an approach to AI auditing that is both interpretable and effective, offering a potential path forward despite the challenges posed by the complexity of modern AI systems.

%% file: sections/intro.tex
\section{Introduction}
\label{sec:intro}

\emph{Auditing} is the systematic evaluation of a system,
often to determine whether it satisfies a predetermined set of criteria. 
With the proliferation of artificial intelligence (AI),
auditing will serve as a vital tool for AI oversight and accountability. 
For example, 
without the  ability to systematically test and assess---or \emph{audit}---for compliance, 
 AI regulations are impossible to enforce. 
Beyond compliance testing, 
auditing also plays several important roles. 
Perhaps most fundamentally, 
it allows for the independent evaluation of developer claims that would otherwise go unverified. 
It can also be used to certify whether an AI technology meets industry standards (e.g., privacy standards) that matter to downstream users (e.g., customers) even when they are not legally required. 
In this way, auditing not only plays an important role in AI accountability, 
but also takes an important step toward developing trustworthy AI. 

Consider, by analogy, the U.S. car industry, 
in which auditing has three important functions.
In the U.S., 
vehicles must adhere to a variety of federal standards and regulations related to safety and emissions, 
which are enforced through audits conducted by the National Highway Traffic Safety Administration (NHTSA). 
Beyond compliance testing, 
car manufacturers are required to disclose information about their vehicles, 
such as their fuel economy (i.e., mileage per gallon), 
which are both internally verified and subject to external audits by the Environmental Protection Agency (EPA). 
To gain an edge over competitors, car manufacturers also make claims about their vehicles;
external and third-party audits validate these claims,
legitimizing them and establishing trust between consumers and manufacturers. 

There is a growing consensus that the AI industry would benefit from similar auditing mechanisms \citep{human-rights-implications-2018, lit-review-audits-2021, outsider-oversight-2022, anatomy-audit-2023}. 
For instance, 
in the European Union (EU), 
the AI Act mandates a mix of internal and third-party audits in the form of ``conformity assessments''  before the release of an AI system or after a substantial modification \citep{conformity-assessment-2024}; %
the General Data Protection Regulation (GDPR) calls for internal audits in the form of impact assessments conducted before data processing \citep{gdpr-impact-assessment-2016};
and the Digital Services Act (DSA) requires annual internal and external audits of risks \citep{dsa-overview-2022}. 
Moreover, to ensure compliance, 
regulatory bodies (e.g., under GDPR, data protection authorities in each member state) are granted broad authority to conduct compliance tests.
There are even provisions (e.g., in the DSA) that grant researchers special access to data and systems so that they can audit for criteria that the regulatory bodies may not consider \citep{dsa-overview-2022}.

Creating a healthy AI auditing ecosystem includes various considerations, 
such as who conducts the audits, who audits the auditors, what auditors test for, whether audits are prospective or retrospective, how often audits are conducted, and more.
For many of these considerations, 
we may be able to look to other industries in which auditing practices are well established for inspiration.
However,
one question is particularly salient for AI auditing:
\begin{quote}
	\begin{center}
	\emph{What type of access and how much access is  needed to conduct a meaningful AI audit?}
	\end{center}
\end{quote}
All auditing procedures require some form of access to the AI system, 
but the particular form of this access must balance (at times, competing) interests:
(1) the protection of intellectual property and proprietary information;
(2) security and privacy concerns;
and (3) resource constraints.
First, the protection of proprietary technologies and data (in particular, of trade secrets) is a key concern for companies, and access granted to auditors is therefore typically limited and carefully controlled. 
For example, EU regulatory bodies must adhere to strict principles of ``necessity'' and ``proportionality'' when handling personal data that is protected by GDPR \citep{gdpr-impact-assessment-2016, gdpr-dpia-2020}.
Second, 
requiring that AI companies open up their technology and datasets can create security and privacy risks. 
Third,  auditors operate with limited resources
(e.g., in budget and technical expertise)
and are therefore interested in the amount of access that allows them to efficiently conduct effective audits. 
Motivated by this question, this work makes the following contributions: 
\begin{itemize}[leftmargin=20pt]
    \item \textbf{Landscape of AI audits.}
    In \Cref{sec: background}, we survey contexts in which AI audits arise, including recent regulatory requirements, and discuss related work. 
    We then examine the various (at times, competing) interests that influence the design and execution of AI audits, 
    motivating our attention to audit \emph{access}.
    In \Cref{sec: aedt case study}, we ground our discussion of audit access and implementation through a case study on New York City Local Law 144. 
    
    \item \textbf{Audit access.}    
    Auditing AI necessarily requires granting the auditor some form of \emph{access} to the underlying system.
    In \Cref{sec:black_box}, 
    we discuss the relative merits and limitations of different types of access to an AI system. 
    While the appropriate form of access is  context-dependent, we conclude that black-box access---the ability to observe a system's behavior on queries of the auditor's choice without examining the system's inner workings---provides the greatest amount of flexibility with the least amount of ambiguity.\footnote{Black-box access is akin to how most users interact with Google Search and ChatGPT. In both cases, a user submits a query (or input), to which they receive a response (or output). This constitutes black-box access in that one collects input-output pairs without knowledge of the underlying mechanism.} We consequently argue that, at minimum, black-box access should be granted to auditors.
    In particular, evidence obtained through black-box access can be definitively attributed to the AI system whereas evidence obtained through other forms of access is less conclusive. 
    Black-box access offers several additional benefits: 
    (1) it does not require direct access to proprietary algorithms or data, i.e., it does not ``open the box;''
    (2) it is agnostic to the underlying AI mechanisms, 
    meaning that the audit does not need to be updated even if the underlying training pipeline changes, i.e., a \emph{standardized} audit can be used across different algorithms;
    (3) it is less resource-intensive than alternate auditing options, 
    allowing continual, comprehensive, and scalable auditing; 
    and (4) it can be run prospectively.\footnote{By ``prospective,'' we mean that (i) the audit can be run before an AI system is deployed and (ii) the AI system can be tested on hypothetical examples, such as extreme scenarios that need not have occurred.}

    \item \textbf{Connections between auditing and hypothesis testing.}
    While we argue that black-box access is minimally necessary to conduct meaningful audits, 
    it is not always sufficient. 
    In \Cref{sec: hypothesis testing},
    we discuss how auditing can be operationalized using hypothesis testing---a well studied statistical framework---and how this perspective provides clarity on two key auditing challenges. 
    First, casting an audit as a hypothesis test provides clear guidance on how much access (\emph{in addition} to black-box access) is needed to obtain the evidence required for a meaningful audit.
    That is, once the parameters of the hypothesis test are set,
    determining what evidence is needed for a meaningful test (i.e., audit) becomes a statistical exercise. 
    Second,
    hypothesis testing has clear parallels to legal procedure and thus provides a way to map between complex statistical auditing methods and the law. 
    In particular,
    we discuss how the null hypothesis can be viewed as legal presumption, 
    placing the burden of proof on the party that wishes to falsify this presumption. 
    Moreover, 
    the ``threshold'' that an auditor selects maps to the ``false positive rate'' or ``false negative rate'' in hypothesis testing.
\end{itemize}

\paragraph{Remark.} 
We note that black-box access is not sufficient for every context. 
As we discuss in-depth,
black-box audits have their blind spots: they cannot speak to the intentions of AI developers, determine whether the
developers adopt best practices, or provide insights that may be necessary for accountability mechanisms (e.g., legal
recourse). To address these gaps, it is often necessary to complement black-box access with additional information
from other sources (e.g., access to related open-source models) as well as gray-box access (e.g., access to log probabilities).
However, full access to the entire learning pipeline is often unnecessary, as scalable auditing procedures
cannot be tailored to the individual features of every AI system.

\paragraph{Remark.} 
The authors are, by training, computer scientists.  Our hope is to provide a careful analysis of how auditing could be implemented in practice given the rapid pace of AI development, 
with a particular focus on the ``access'' question.  In an effort to provide an interdisciplinary discussion on AI auditing, this piece is a blend of styles. Our primary goal is to connect tools familiar to technologists with concepts that resonate with regulators;  namely, hypothesis testing, evidence gathering, and legal presumption. This piece is therefore intended for two audiences:  (i) policymakers interested in the implementation of AI audits, and (ii) computer scientists interested in developing audits.

%% file: sections/related_work.tex
\section{Background and terminology}
\label{sec: background}

In this section, 
we briefly review auditing and its relation to AI. 
Of particular note is \Cref{table:legislation_examples}, 
which summarizes recent AI audit requirements. We note that AI auditing is a broad topic spanning computer science, law, and the social sciences, and we provide a more extensive discussion of related work in \Cref{sec:ai_audit_review,sec: additional related}.

\subsection{A brief introduction to auditing}

An audit is a systematic assessment of an organization, system, and/or process. 
Audits are conducted for many reasons, including (i) testing for \emph{compliance with regulations}, (ii) determining whether a technology meets \emph{certification standards}, (iii) validating \emph{claims made by system designers}, (iv) monitoring an organization's \emph{internal practices}, and
(v) uncovering \emph{vulnerabilities}.
As examples,
the 2002 Sarbanes-Oxley Act made financial auditing commonplace as a tool to detect fraud and confirm the accuracy and completeness of financial reports; 
to earn a fair trade certification, vendors regularly undergo audits to ensure they uphold fair trade practices; 
many organizations audit themselves to detect, e.g., financial waste;
and software designers often undertake audits to discover security vulnerabilities.

Generally, there are three types of auditors: internal, external, and third-party auditors. An \textbf{internal} auditor is selected from within an organization that seeks to audit itself. An \textbf{external} auditor is an independent party that is hired by the organization (e.g., its board of directors) to perform an audit. External auditors typically enter into a contractual agreement with the organization that outlines, for instance, the scope of the audit and level of engagement. 
Like an external auditor, a \textbf{third-party}  auditor is not a part of the organization being audited. Unlike external auditors, third-party auditors are not hired by the organization that they audit. For example, a third-party auditor may be a regulatory agency or be employed by a regulatory agency. 
Third-party auditors also include journalists and non-profits (as long as they are not hired by the organization being audited).
In order to ensure that auditors provide objective and high-quality reports, \emph{auditors are also subject to audits}. 
It is customary for auditors to audit one another and, in certain industries, auditors are overseen by government agencies, such as the Public Company Accounting Oversight Board (PCAOB) in the US and the Financial Reporting Council (FRC) in the UK. 

Audits can be initiated at different times and run with varying frequency. 
\textbf{Retrospective audits} evaluate a system's past behavior. 
For example, 
audits that examine financial records or system performance are retrospective. 
These retrospective audits can be ad hoc (in response to specific events), continuous, or periodic. 
In contrast, \textbf{prospective audits} are proactive. They can either (i) be performed before deployment or (ii) be tested on examples/contexts that have not yet arisen.
As an example of (i), prospective audits can be run prior to an AI system's release or after major modifications to evaluate potential risk. 
As an example of (ii), 
prospective audits can also assess how a system %
\emph{would} behave under conditions that have not yet occurred, e.g., in extreme scenarios.

Although out of the scope of this work, 
a final consideration for auditing is the development of relevant \textbf{metrics}, \textbf{measurement methods}, and \textbf{standards},
which has become a central focus of the U.S. National Institute for Standards and Technology (NIST) and European Commission. 
These efforts center around \emph{what} AI audits should test for, which we take to be predetermined in this work in order to focus on AI auditing's implementation challenges.

\subsection{AI auditing}

\paragraph{Legally required AI audits.}
While some organizations audit themselves
and some organizations are audited by journalists, researchers, and more, 
there have historically been few instances of legally mandated AI audits. 
That is beginning to change; \Cref{table:legislation_examples} provides a (non-exhaustive) list of auditing requirements for
the European Union (EU) Artificial Intelligence (AI) Act, 
the EU General Data Protection Regulation (GDPR), 
the California Consumer Privacy Act (CCPA), 
the US Algorithmic Accountability Act (AAA), 
the New York City (NYC) Local Law 144, 
Canada's Directive on Automated Decision-Making (CDADM), 
and the EU Digital Services Act (DSA).

\paragraph{Laws that indirectly mandate AI audits.}
In addition to those detailed in \Cref{table:legislation_examples}, there are several domains in which AI audits are indirectly required. 
For example, the Dodd-Frank Wall Street Reform and Consumer Protection Act and Sarbanes-Oxley Act (SOX) both require audits to check for compliance with financial regulations.
Although neither explicitly mention AI or algorithms, 
they have both become common tools in organizations that are required to comply with Dodd-Frank and SOX.
As a result, they indirectly fall within the scope of these required audits. 

\paragraph{Other contexts.}
Audits are also a common tool for ensuring compliance with the law, 
even when the audits  are not explicitly mentioned.
Regulatory bodies and third-parties will often audit in order to hold organizations legally or publicly accountable, e.g., the Children’s Online Privacy Protection Act (COPPA) does not explicitly mention audits, but the Federal Trade Commission (FTC) conducts investigations to verify compliance with COPPA's mandates \cite{FTC2022EdTechPolicy}.

\input{sections/abbrev_table}

Academic researchers, companies, and journalists may also conduct audits that fall outside the scope of compliance auditing. These audits are often used to verify that AI satisfies \emph{normative} standards and meets developers' claimed specifications, even if these standards are not (or not yet) legally mandated. Some of these audits are an extension of longstanding approaches to AI and ML \emph{evaluation}, which has historically focused on measures of performance (e.g., popular benchmark datasets like those in the UCI Machine Learning Repository \cite{kelly2024uci} or ImageNet \cite{deng2009imagenet}, and standardized evaluation suites like the Holistic Evaluation of Language Models (HELM) \cite{Liang2023HolisticEO}, among many others). 
Others have advocated for audits that look beyond performance, toward considerations such as bias, discrimination, and equity \cite{methods-discrimination-2014}.
Several well known audit studies, such as those that brought light to potential discrimination in criminal justice algorithms \cite{angwin2016machine} and facial recognition algorithms \cite{Buolamwini2018GenderSI}, underscore the need for such audits. 

Efforts to audit for these kinds of harms are rapidly evolving, particularly in the context of newer generative AI technologies, but some common practices have started to emerge. These include ``sock puppet audits," in which auditors programmatically impersonate users of an online platform to investigate the platform's algorithmic behavior \cite{methods-discrimination-2014, Hannk2014MeasuringPD, personalization-search-2017, Haroon2023AuditingYR, Hosseinmardi2023CausallyET}, and ``sociotechnical audits,'' in which auditors explicitly study interactions between humans and algorithms to contextualize an algorithm's behavior relative to the social environment in which it is deployed \cite{sociotechnical-audits-2023}. \citet{ai-accountability-gap-2020} propose an end-to-end framework for {internal} audits, e.g., those which are voluntarily conducted by a company or model provider, which nevertheless provides guidance on audits more broadly.

Auditing frameworks and methodology have also emerged for more specific contexts, including the auditing of social media platforms \cite{regulating-filtering-2020, framework-regulating-social-media-2023}, auditing for (a particular notion of) fairness via active learning \cite{active-fairness-2022}, and auditing for differential privacy \cite{2002.04049, 2006.07709, 2302.07956}. Red teaming, a classical approach to ensuring the security of software systems, has emerged as a popular method for evaluating modern generative AI systems (e.g., large language models) \cite{Perez2022RedTL, Ganguli2022RedTL, Rando2022RedTeamingTS, Yu2023GPTFUZZERRT}, though questions remain about its efficacy in this context \cite{2401.15897}. We provide additional background on other forms of AI auditing in \Cref{sec:ai_audit_review}, and a discussion of other related work in \Cref{sec: additional related}.

\paragraph{Limitations on audit access}
Auditors are typically limited in their ability to access the systems  they audit. 
In some cases, 
limited access is mandated by the law, 
e.g., 
the CCPA states that ``nothing in this section [on risk assessments] shall require a business to divulge trade secrets;'' 
 GDPR similarly states that data processing measures (including audits) ``should be appropriate, necessary and proportionate in view of ensuring compliance with this Regulation;''
 and the DSA states that audits ``shall ensure an adequate level of confidentiality and professional secrecy in respect of the information obtained.''
In addition to trade-secret and privacy concerns, 
auditor access may be limited for cybersecurity reasons. 
At times,
third-party auditors do not notify (and therefore do not cooperate with)  the organization being audited, and thus have minimal access to the AI system. 
This consideration of \emph{access} will be central to our discussion.

%% file: sections/abbrev_table.tex
{
\onecolumn
\fontsize{9pt}{9pt}\selectfont
\begin{longtable}{L{0.53in} L{0.85in} L{1in} L{2in} L{0.88in}}
\caption{Examples of legislation that require audits of data-driven or AI algorithms}
\label{table:legislation_examples}\\
\toprule

\textbf{Law} & \textbf{Enforced by} & \textbf{Performed by} & \textbf{Audit frequency and requirements} & \textbf{Penalty} \cr
\midrule
\\[-10pt]
EU GDPR (2016) & Data Protection Authorities (DPAs) in EU member states. Overseen by European Data Protection Board (EDPB). & Data controllers (typically internal), potentially with the help of third parties and data processors. 
& 
Mandates Data Protection Impact Assessments (DPIAs): descriptions of data processing, purposes, risks to rights \& freedoms of subjects, measures to address risks, as laid out in Article 35. DPIAs are required before high-risk data processing and when there is a change of the risk represented by processing. 
& Up to €10M or 2\% of annual worldwide turnover (whichever is higher). Up to €20M or 4\% of annual worldwide turnover (whichever is higher) for severe violations.  \cr
\hline
\\[-10pt]
EU AI Act (2023) & National competent authorities in EU member states. Overseen by European Commission (EC). &
AI system providers (internal) or notified bodies (third-party), depending on the existence of harmonized standards or common specifications.
& High-risk AI systems must undergo conformity assessments to ensure they meet requirements for safety, transparency, human oversight, data, and more. Requires assessment before system is on the market, ongoing post-market monitoring, and whenever system is substantially modified. &
Determined by member states; Some infringements up to €30M or 6\% of annual worldwide turnover, whichever is higher.\cr
\hline
\\[-10pt]
CCPA (2018) & California Attorney General and California Privacy Protection Agency. & 
Businesses whose data processing presents significant risks to consumer privacy or security.
& 

    When business' processing of personal information poses 
    significant risk to consumers’ privacy or security,
    (i) must perform a cybersecurity audit on an annual basis that determines the data processing security risk, and 
    (ii) must submit a risk assessment on a     ``regular'' basis that weigh the benefits of processing personal information against potential risks to consumer rights. 
    States that nothing in (ii) shall require a business to divulge trade secrets.

    & Up to \$7.5K per intentional violation; additional penalties given by California Privacy Protection Agency. \cr
\hline
\\[-10pt]
US AAA (2023${}^\dagger$) & Federal Trade Commission (FTC) or, as necessary, the attorney general of a state. & Covered entities over which the FTC has jurisdiction, deploys any automated decision system, and meets other stated criteria.
& 
Requires ongoing impact assessments of automated decision systems or decision processes that use them to make critical decisions. Businesses must consult internal and external stakeholders.  
Mandates annual reports submitted to the FTC and initial summary report prior to deployment. 
& 
Determined by the FTC. \cr
\hline
\\[-10pt]
NYC 144 (2021) & NYC Dept. of Consumer \& Worker Protection (DCWP) & 
Independent auditor (external) & 
Requires bias audit (impartial evaluation) that tests whether automated employment decision tool's disparate impact on persons of any ``component 1 category'' to be conducted annually and prior to first use. A summary must be made publicly available. Conducted prior to first use and annually.
& Up to \$1.5K per instance; others determined by enforcement body.\cr
\hline
\\[-10pt]
CDADM (2019) &
Treasury Board of Canada Secretariat (TBS), potentially alongside other oversight bodies.
& Federal institutions using production automated decision systems. 
& 
Requires algorithmic impact assessments (AIAs) to assess and reduce risk. AIA is currently a questionnaire composed of 51 risk and 34 mitigation questions.
AIAs are required prior to production of automated decision system; 
to be reviewed and updated on a scheduled basis, including when system undergoes changes. 
There are also testing and monitoring requirements, e.g., on unintended biases, unfair impact, and for unintentional outcomes. 

& Case-by-case, as determined by TBS and potentially other oversight bodies. \cr
\hline
\\[-10pt]
EU DSA (2022) & Digital Service Coordinator (DSC) in each EU member state and the EC. %
& Audits to be performed by independent auditor (external), with some guidelines (e.g., cannot audit >$10$ consecutive years). 
Risk assessments and ongoing monitoring to be conducted internally. %
&  Requires independent audits of providers of very large online platforms and of very large online search engines that test compliance with the obligations set out in Chapter III of the DSA to be conducted annually. Also requires that they perform assessments of systemic risks and continuous monitoring of risk mitigation strategies. 
&
Up to 6\% of annual worldwide turnover for failure to comply; periodic penalties must not exceed 5\% of average daily worldwide turnover or income per day.

\\
\bottomrule\\[-10pt]
\multicolumn{3}{L{2.6in}}{\scriptsize $\dagger$ Proposed but not passed}  \\
\end{longtable}
}

%% file: sections/aedt_case_study.tex
\section{Illustrating AI auditing challenges}
\label{sec: aedt case study}

To highlight the challenges of implementing AI audits, we will ground our discussion in a case study of New York City (NYC) Local Law 144, which we will return to throughout our discussion. 

\paragraph{NYC Local Law 144.}
This law, which was enacted in 2021 and took effect in 2023, governs the use of automated employment decision tools (AEDTs). 
It ``prohibits employers and employment agencies from using an automated employment decision tool unless the tool has been subject to a bias audit within one year of the use of the tool, information about the bias audit is publicly available, and certain notices have been provided to employees or job candidates'' \cite{dcwp}.
The law defines a bias audit as  an ``impartial evaluation by an independent auditor'' that assesses  ``the tool's disparate impact'' with respect to  job category, sex, and race or ethnicity;
and an AEDT is defined as any ``computational process'' that is ``derived machine learning, statistical modeling, data analytics, or artificial intelligence'' that substantially supports, assists, or replaces any ``discretionary decision-making'' task in hiring or promotion, e.g., by providing an assessment of a candidate's skills or screening candidates for interviews \cite{nyc_code}.

Although the administrative code enacted in 2021 does not specify implementation details, rules released by city agencies in 2023 provided more specific guidance that we discuss below \cite{nyc_rules}. 

\paragraph{Regulations providing guidelines for NYC Local Law 144.}
The rules released in 2023 provide greater details on what should be reported in the audits and what data should be used. 
It specifies that a bias audit must, at minimum, calculate selection rates, scoring rates, median scores, and impact ratios for race/ethnicity, sex, and intersectional categories.
The audit may also interrogate other features of the AEDT, though no further assessment is required.

The rules also require that bias audits be conducted on ``\emph{historical data}'' (data collected during the employer's or employment agency's use of the AEDT or, in some circumstances, others' use of an AEDT). 
This audit must be performed on sufficient historical data to be ``\emph{statistically significant,}'' though the rules do not clarify key elements of this requirement, which can lead to two very different outcomes (see \Cref{sec: hypothesis testing}).
The law also makes allowances for the use of ``\emph{test data}'' (any data that is not historical data, such as synthetic inputs) only if sufficient historical data is unavailable to conduct a statistically significant audit, though it does not provide explicit requirements or guidance on how such data should be acquired or generated.
It does require that a ``summary of results of the bias audit must explain why historical data was not used and describe how the test data used was generated and obtained'' \citep{nyc_rules}. Similarly, the law does not restrict the ability to exclude certain historical data (e.g., specific time periods) from the audit, but requires that these choices be publicly disclosed. 
 
 Interestingly, \emph{imputed} and \emph{inferred} data cannot be used in a bias audit, implying that if race, ethnicity, or gender is not reported for an individual, 
 then they would be excluded from the corresponding bias analysis, even if the AEDT imputes or infers race, ethnicity, or gender from the individual's other attributes. Thus, even if the historical dataset contains many samples, statistically significant results may only be possible with the help of synthetic data if candidates' demographics were not collected by the employer/employment agency.
 Several other elements of the bias audit remain unspecified, including
whether the bias audit is performed over the AEDT alone, over the entire decision-making process that includes the AEDT, or over a subset of it.

\paragraph{Open implementation challenges.}
The rules discussed above provide clarifications on NYC Local Law 144 but also defer many implementation details to the auditor (or relevant standards-setting body), including the manner in which the auditor \emph{accesses} the algorithm, \emph{what} evidence they should collect, \emph{how} one should interpret that evidence. 

For example, there are many cases in which test data may be required, as described in the rules (e.g., if there is insufficient historical data to conduct a statistical test, or demographic information is missing). 
However, there are multiple ways in which the auditor could construct the test data, each of which has the potential to yield very different audit outcomes.
As another example of ambiguity in audit implementation,
the rules require that bias audits yield statistically significant results, but what model should the auditor use to assess ``statistical significance?'' 
Both issues are implicitly related to the ``access question'' that motivates our discussion: 
creating test data requires, at a minimum, query (or black-box) \emph{access} to a system, and constructing a reasonable model similarly requires insights into the AEDT (which can be viewed as a form of \emph{access} to the training procedure, 
as we discuss extensively in \Cref{sec:black_box}).
These challenges are closely linked to unresolved questions around audit reporting, 
such as how the auditor should communicate the results of (often highly technical) audits.
Thus, an examination of NYC Local Law 144 highlights broader challenges around \emph{access}, \emph{evidence}, and \emph{interpretation} that arise in the operationalization of AI auditing more broadly.

\paragraph{Connections to this work.}
As the prior section highlights, these concerns are broadly relevant for AI audits, thus motivating our focus, in \Cref{sec:black_box}, on the \emph{type} of access an auditor needs to audit AI systems and, in \Cref{sec: hypothesis testing}, on a principled and flexible framework for (1) determining what \emph{additional} assumptions or information are necessary to conduct an AI audit and (2) gathering and interpreting evidence gathered from the audit. 
 In \Cref{sec: hypothesis testing}, we discuss what kind of access to an AI system is necessary to perform a meaningful audit. We survey possible options and argue that the ability to \emph{query} an AI system---construct arbitrary inputs and observe the system's outputs---is almost always necessary (i.e., minimally required) to perform an informative audit. In other words, auditors should  at least be granted \emph{black-box} access to an AI system. We proceed to connect AI auditing to the field of hypothesis testing in \Cref{sec: hypothesis testing}, which we find provides a principled framework for collecting and interpreting evidence in the context of an audit. 
 We show how hypothesis testing as a framework for AI audits can (i) inform what type of access (in addition to black-box access) an auditor may require, (ii) clarify the settings (specifically, the model and assumptions) under which an AI audit produces meaningful results, and (iii) produce alignment between  AI audit methods and legal procedure due to parallels between hypothesis testing and legal concepts.

%% file: sections/black_box_2.tex
\section{Types of auditing access}\label{sec:black_box}

In this section, we address our motivating question: what form of access does an auditor require to perform a meaningful audit?
In practice, answering this question is not straightforward due to competing concerns: while the auditor necessarily requires some form of access to the underlying system to conduct an audit, it is often desirable to keep this access ``minimal'' to avoid unnecessarily exposing proprietary technologies, compromising private data, and introducing cybersecurity risks, among other concerns.
To address this question, we discuss the relative merits of four types of auditor access: access to the training data, the training procedure, the model architecture, and 
(white- and black-box access) to the trained model. 
 Our discussion below hews closely to that of \citet{auditing-blog}.

\subsection{Option 1: Access to training data}\label{sec:option1}

AI models learn patterns and relationships that are exhibited in their \emph{training data}. Because this data is one of the key determinants of an AI system's behavior, an auditor may wish to examine it for suggestive evidence of potential harms or failure modes.
For example, over- or under-representation of a population in the training data can lead to bias \citep{fairness-frontiers-2018}. Similarly, 
differences between the test and training data distributions can lead to generalization failures  \citep{domain-gen-survey-2021}.
Nonetheless, access to the training data alone is typically insufficient for a rigorous audit. 
The primary reason is that the same training data can induce many different downstream models, whose behavior ultimately depends on the entire training pipeline (e.g., the choice of hyperparameters, model architecture, and learning algorithm). Although they arise from the same training data, these models may differ substantially along nearly any dimension of interest, including accuracy, fairness, and robustness. This ``model multiplicity'' \citep{multiplicity-2022} or ``underspecification'' \citep{underspecification-2022} is an unavoidable feature of modern AI systems.
As such, it is not only difficult, but often impossible for an auditor to conclusively characterize a system's behavior from its training data alone. 

Nonetheless, examining the training data can play a key role in AI auditing.
For instance, 
auditing a company's data acquisition, cleaning, balancing, privacy, and provenance practices can encourage good data hygiene. 
Although requiring that the training data meet a strict set of property requirements does not generally have the intended effect (most bright-line rules are easily circumvented due to the underspecification phenomenon mentioned above), 
\emph{data disclosure audits} can encourage responsible developer practices and prevent downstream harms \cite{model-cards-2019, pushkarna2022data, gebru2021datasheets}. 
For example, in the context of modern large language models (LLMs), the common practice of publishing a ``knowledge cutoff date"---roughly, the most recent date of the data on which the LLM was trained---may obscure important heterogeneity in the model's temporal understanding across different tasks; disclosure audits may therefore encourage developers to release finer grained or resource-specific knowledge cutoffs that could enable more informed use \cite{2403.12958}.

\subsection{Option 2: Access to training procedures}

Another option is to grant an auditor access to the \emph{training procedure},
which includes details such as the general model class (e.g., decision trees, linear models, transformers), objective function used to optimize parameters, hyperparameter tuning methodology, quality checks, or model selection criteria. The training procedure can be viewed as a roadmap for how the AI system was developed. For example, around 2017, one of the objective functions in Facebook's ``News Feed`` algorithm placed five times more weight on ``reactions'' than it did on ``likes'' when ranking content in a user's feed \citep{fb-formula-2021}. This change had the unintended consequence of amplifying emotional content; indeed, the company's own data scientists found that posts that ``sparked [the] angry reaction emoji were disproportionately likely to include misinformation, toxicity and low-quality news" \citep{fb-formula-2021}. These issues were not detected by Facebook until 2019 and only made public in 2021 by a whistleblower \citep{fb-formula-2021}. An earlier audit of the training procedure, including the objective functions, may have encouraged the company to scrutinize and justify (and, if appropriate, abandon) these design choices. 

However, as with the training data, a particular training procedure can yield many downstream models, and there is no guarantee that these models behave similarly.
The resulting model depends on various other factors, including the training data, model weights at initialization, and more. Therefore, while an auditor who is given access to a system's training procedure can perform sanity checks, they typically cannot draw precise conclusions about the system's ultimate behavior. Furthermore, since the training procedure lays out the steps taken to produce the AI system, access to training procedures is carefully guarded by companies;
of the forms of access discussed, the training procedure is arguably the most valuable intellectual property associated with a commercial product.

\subsection{Option 3: Access to the model skeleton}

The next form of access we consider is access to the model ``skeleton", which we also refer to as the ``untrained" model. 
This ``skeleton" refers to the specific model \emph{class} that is used (e.g., the specific neural network architecture) and the way that different system components interact (e.g., how an AI model interacts with other models within the same system). 
 Importantly, this skeleton is disclosed without the model parameters (i.e., weights).

One defining feature of this form of access is that it reveals the key interfaces of an AI system. 
From the model skeleton, 
an auditor can determine the expected inputs (e.g., types of features) and outputs (e.g., a floating point number between 0 and 1) of the model. 
The auditor can also ascertain how many components make up the AI system and 
the relationship between different components of the AI system. 
For example, 
suppose that a job applicant's information and resume are first sorted into one of several job categories,
processed by appropriate algorithms, 
before being assigned a score between 0 and 1, 
which is finally thresholded to produce a hiring recommendation. 
Then, this entire ``pipeline'' would be captured by the model skeleton. 
In this way, access to the model skeleton provides perhaps the most \emph{interpretable} view of an AI system. 
Indeed, when the social media platform X (formerly known as Twitter) voluntarily released their recommendation model skeleton, \citep{twitter-algo}, it revealed qualitative insights into X's content curation algorithm. 
For example, the public could glean that X ``sources half of a user’s content from in-network tweets (i.e., from accounts that the user follows) and the other half from out-of-network tweets" \citep{auditing-blog}.

As with the training procedure, 
a model skeleton allows the auditor to perform sanity checks to flag obvious flaws. 
It can even be used to identify discrepancies between an AI company's claims and the deployed system. 
The model skeleton alone, however, is not enough to characterize the precise behavior of an AI system. 
As mentioned in Section \ref{sec:option1}, 
models with the exact same skeleton can behave very differently, making it difficult to verify whether an AI system complies with a specific rule or meets a given standard from access to the model skeleton alone. 

\subsection{Option 4: Access to the trained model}\label{sec:trained_model_access}

We now turn to the last option we consider: access to the final trained model.
In contrast to the other three forms of access, access to the trained model allows the auditor to inspect the \emph{specific} model that is or will be released. 

There are several different flavors of access to the trained model.
\emph{Black-box access} allows the auditor to query the model on the auditor's choice of inputs and observe the outputs but nothing more, yielding a series of input-output pairs. 
For example, sending ChatGPT queries (or ``prompts'') and observing how ChatGPT responds is a form of black-box access. Similarly, observing the ratings that a hiring algorithm assigns to job applicants is also a form of black-box access. 
On the other hand, \emph{model-weight access} allows the auditor to not only query the model, but also to see the entire trained model, including the trained model parameters (i.e., the model weights). 
By analogy, one can think of black-box access as an auditor being able to crash-test a car, whereas an auditor with model-weight access would also be able to inspect every component of the car. 
In between these two, there are other forms of access that have been explored, including \emph{fine-tuning access} (the ability to fine-tune the final trained model on a dataset of interest) and \emph{log-probabilities access} (the ability to view ``probabilities'' the model assigns to different prediction outcomes, before the final prediction is produced in most neural networks). 
In this way, there are tiers of access to the trained model. For instance, model-weight access is strictly stronger than log-probabilities access (in that one can always obtain log probabilities from model weights), which is stronger than black-box access.

Of the four options considered in this section, 
only access to the trained model circumvents the problem of ``underspecification" or ``model multiplicity" \cite{underspecification-2022, multiplicity-2022}. 
This implies that \emph{access to the trained model is minimally necessary for meaningful audits} because it best indicates how the system behaves when deployed.
However, there are limitations to auditing the trained model alone. 
An auditor cannot gain insight into the developers' process and reasoning from the trained model alone, 
which is not always satisfactory from a broader accountability perspective. 
Much of US law, for instance, considers \emph{intent} {when determining culpability}. 
Moreover, auditing developers' process and reasoning can encourage safer and more thoughtful industry practices. 
As such, further access is often needed.

\subsection{Key takeaway: Black-box access is minimally necessary}

The four types of access described above are not exhaustive. 
For example, \emph{white-box access} allows the auditor to view the model weights, training procedure, and source code but typically not the training data.
Although there are numerous access options, 
one can view them as follows: putting resource constraints aside, the \emph{stronger} the form of access, the closer the auditor is to being able to \emph{reproduce} the AI system from scratch.
Therefore, granting auditors access which is overly broad can introduce its own risks, such as undermining a company’s competitive advantage or allowing bad actors to exploit system vulnerabilities.

We argue that black-box access---the weakest form of access to the trained model---should be minimally required, as it provides concrete evidence about the specific AI system being audited (see \Cref{sec:trained_model_access}). In addition, it has several characteristics that may address implementation difficulties that auditors face, as enumerated below:
\begin{enumerate}
	\item \textbf{Minimal access.}
		Black-box access only allows the auditor to view the outputs of an AI system. It thus allows the auditor to assess the system that is ultimately deployed without further insight into the underlying implementation, which can allay concerns about revealing trade secrets, compromising data privacy, or exposing security vulnerabilities.
		
	\item \textbf{Prospective.}
	Although this property is shared with model-weight  and log-probabilities access (but not all flavors of trained-model access), black-box access gives the auditor the ability to prospectively inspect the AI system. 
	That is, the auditor can test the AI system on inputs of their choice and observe how the system behaves in situations that may not have arisen. 
	This allows the auditor to perform pre-deployment assessments, 
	and it can also be used to stress-test the system under extreme scenarios that have not yet arisen in natural datasets.
	
	\item
	\textbf{Model agnostic.}
	 Because black-box audits do not look ``under the hood,'' they are agnostic to the inner workings of the AI system.
	Perhaps the key benefit is that
	the audit does not need to be adapted to the underlying AI system, 
	making it possible to develop a \emph{standardized} audit procedure that works across multiple AI systems. 
	For example, even when developers change their model architecture, an auditor does not need to devise a new black-box audit
	as long as the input-output ``types'' are consistent.
	Another benefit of model agnosticism is that it allows the auditor to test how the model behaves end-to-end without necessarily requiring that the auditor be technically proficient (which is often needed if an auditor wishes to conduct model-specific audits).

	\item 
	\textbf{Well suited to AI.}
	Finally, it is worth remarking that AI systems are well-suited to black-box audits.
	Consider, for example, auditing a firm's (non-AI) hiring practices for discrimination. A black-box audit would require gathering everyone who plays a role in hiring, providing them with a set of applications,
	asking them to evaluate each as they normally would, and observing their decisions. 
	This process is not scalable, as it would require auditors and the firm to invest significant time and resources. Perhaps more importantly,
	those involved with the audit can easily manipulate the outcome by misreporting their true preferences on who to hire. On the other hand, 
	it is straightforward to repeatedly query an AI system, and the results of black-box AI audits are guaranteed to be faithful to how the AI system would behave in practice.
\end{enumerate}
As discussed in this section, no form of access is uniformly better than its alternatives, and the appropriate form of access is context-dependent.
We argue however that black-box access is minimally necessary for an informative audit, reduces some of the risks associated with stronger forms of access, and offers other advantages described above, begging the question: if black-box access is minimally necessary, \emph{in what contexts is stronger access needed, and how should one develop standardized audit protocols based on the available level of access?} We examine this question next.

%% file: sections/framework.tex
\section{Hypothesis testing and its connections to audits}
\label{sec: hypothesis testing}

In the prior section, we discuss different forms of access to an AI system, and argue that {black-box} access is often necessary to conduct an informative audit.
When, however, does the auditor need more than black-box access, and how should the auditor interpret the evidence that they gather? 
In this section, we draw parallels between auditing and hypothesis testing, 
using NYC Local Law 144 (as described in \Cref{sec: aedt case study}) as an illustration. 
We show that this connection helps (i) clarify how an auditor can interpret and communicate the results of an audit, and (ii) produce precise guidance on how much access an auditor needs. 
We highlight that the null hypothesis can be viewed as a formalization of the relevant legal presumption, 
and the evidence needed to reject the null hypothesis can be viewed as the corresponding burden of proof. 
We provide a discussion of related work in hypothesis testing  in \Cref{sec:ai_audit_review}.

\subsection{Setup}\label{sec:setup}

Consider a model developer or operator, 
who we refer to as the \emph{AI provider} for the remainder of this section. 
The provider employs an algorithm $f \in \cF$, 
where $\cF$ is a class of mappings from values in $\cX$ to distributions over a countable  set $\cY$ as denoted by $\Delta(\mathcal{Y})$.
For example, in the context of hiring decisions, 
$f$ could map an applicant's characteristics $x \in \cX$ to a prediction $f(x)\in \cY \subset [0, 1]$ of the applicant's fit for a given role. 
Let $p_x$ denote the true (possibly unknown) marginal distribution of $x$. 
The auditor is interested in determining whether the provider's algorithm $f$ complies with a requirement of interest. We denote this requirement by $g: \cF \rightarrow \reals$. 

\begin{definition}\label{def:g-compliance}
    We say that an algorithm $f \in \cF$ is $g$-\emph{compliant} if and only if $g(f) \le 0$.    
\end{definition}
When the property $g$ is clear from context, we simply say an algorithm is \emph{compliant}. In a \emph{black-box audit},
the auditor has access to $N$ input-output pairs $(x_i, f(x_i))$.
We denote  the \emph{evidence} that an auditor has access to (including that gathered through black-box access)  by $\mathcal{E}$. 
Thus, the auditor's task is as follows: 
\begin{center}
\emph{
    Determine whether $f$ is $g$-compliant given evidence $\mathcal{E}$.
    }
\end{center}

\begin{example}[Maximum loss]\label{ex: max loss}
    Requiring that $f$'s maximum loss $\ell$ over some $S \subseteq \cX$ is at most $\eta$ is equivalent to requiring that $g(f) \leq 0$, where $g(f) = \max_{x \in S} \ell(f(x), x) - \eta$.
    Depending on the definition of loss, 
    one can audit for \underline{minimax fairness} (by defining loss as negative performance), 
    \underline{worst-case harm} (by defining loss as the output's harm, e.g., toxicity level), 
    and even \underline{copyright infringement} (by defining loss as the dissimilarity between $x$ and the copyrighted work).
\end{example}

\begin{example}[Group fairness]\label{ex:group_fairness}
    In the area of algorithmic fairness, 
    group fairness typically reflects a notion of parity across groups. 
    For example, 
    one notion of group fairness known as ``statistical parity'' requires that the rate at which a binary classifier $f: \cX \rightarrow \{0, 1\}$ selects members of group $G_1$ is at most $\eta > 0$ far from the rate at which $f$ selects members of group $G_2$ under some distribution $p_x$ over $\cX$. 
    This is equivalent to requiring that $g(f) \leq 0$, where
    $g(f) = \left|\bbE[ f(x)  \,|\, x_G = G_1 ]  - \bbE[f(x) \,|\, x_G = G_2 ] \right| - \eta.$
    The expectation above is taken over $x \sim p_x$, and
    $x_G \in \{ G_1, G_2 \}$ is the feature in $x$ denoting group membership. 
\end{example}

\begin{example}[Individual fairness]
    Another notion of algorithmic fairness requires that ``similar individuals be treated similarly,'' as captured by the criterion: $D(f(x), f(x')) \leq L d(x, x')$ for all $x, x' \in \cX$;
    distance metrics $D$ and $d$ on $\cY$ and $\cX$, respectively;
    and Lipschitz constant $L > 0$  \citep{fairness-through-awareness-2011}. 
    This is equivalent to requiring that $g(f) \leq 0$, 
    where $g(f) = \max_{x, x' \in \cX} \frac{D( f(x) , f(x') )}{d(x, x')} - L$. 
    
\end{example}

One could similarly cast selection rates or impact ratios (from \Cref{sec: aedt case study}) in this format.

\subsection{Hypothesis testing and  the burden of proof}\label{sec:burden_proof}

Given an algorithm $f$ and auditing criterion $g$, 
the auditor seeks to determine whether $f$ is $g$-compliant using $\mathcal{E}$.
Below, we discuss two possible hypothesis tests before describing the general hypothesis testing procedure in Section \ref{sec:HT_proc}.

\paragraph{Presumption of compliance.}
Consider an auditor who seeks to discern which of the following hypotheses holds:
\begin{align}
    H_0: g(f) \leq 0 , \qquad H_1: g(f) > 0 . \label{eq:HT_compliant}
\end{align}
Let $H \in \{H_0, H_1 \}$ denote the ground-truth state.
For example, 
if $f$ is compliant, then $H = H_0$; 
if $f$ is not, then $H = H_1$.

The auditor does not know $H$ {a priori}, and the auditor's goal is to develop a decision test or \emph{rule} $\hat{H}$ such that
$\hat{H}(\mathcal{E}) = H_0$ if $f$ is compliant and $\hat{H}(\mathcal{E}) = H_1$, otherwise.
The auditor would like $\hat{H}$ to match $H$ for all $f \in \cF$ because this means that the auditor has neither false positives nor false negatives, 
as formalized in Section \ref{sec:HT_proc}.

$H_0$ is known as the \emph{null hypothesis}. 
In practice, 
the implication is that
the auditor's \emph{presumption} under \eqref{eq:HT_compliant} is that $f$ is compliant. 
The auditor therefore assumes (and reports) that $f$ is compliant \emph{unless the evidence allows them to confidently reject this presumption}.
This framework therefore highlights what evidence is needed for an auditor to reject $H_0$ when $H_1$ is indeed the true hypothesis, 
i.e., what information the auditor needs to prove that an AI system is non-compliant when it is indeed non-compliant. 
Though subtle,
\emph{this point is crucial.} 
By viewing auditing as hypothesis testing, 
it becomes clear what it means for the available information to be (in)sufficient for auditing. 
Below, we show that the hypothesis test can be reversed to instead presume non-compliance.

\paragraph{Presumption of non-compliance.}
Consider a different set of hypotheses: 
\begin{align}
    J_0: g(f) > 0 , \qquad J_1: g(f) \leq 0 \label{eq:HT_noncompliant}
\end{align}
Relative to \eqref{eq:HT_compliant}, 
the null and alternate hypotheses have been swapped. 
As before, 
there is a ground-truth state  $J \in \{J_0, J_1 \}$, 
and the auditor's goal is to develop a decision rule $\hat{J}$ such that, given evidence $\mathcal{E}$, 
the decision $\hat{J}$ approximates $J$ well across all $f \in \cF$. 
In this case, the null hypothesis $J_0$  (i.e., the legal presumption) is that the algorithm is not compliant.

\paragraph{Burden of proof: Which  test should the auditor use?}
The null hypothesis reflects the auditor's presumption and, accordingly, who bears the burden of proof. 
For example, NYC Local Law 144 requires bias audits to be ``statistically significant'' but does not specify the null hypothesis, among other necessary modeling assumptions that we discuss further below.
When such audits are required but do not explicitly require the AI provider to disclose any information to auditors,
then,
under \eqref{eq:HT_compliant},
the AI provider is \emph{not incentivized} to disclose information (i.e., to contribute evidence $\mathcal{E}$ to the auditing process). 
To see this, observe that 
since the auditor can only reject the null hypothesis $H_0: g(f) \leq 0$ if they have enough evidence to do so, 
the burden of proof is on the  auditor or governing body. 
Therefore, NYC Local Law 144, as it stands, incentivizes employers and employment agencies to release minimal amounts of historical data (recall that it does not specify the timeframe of the data that must be provided) if the null is \eqref{eq:HT_compliant}.

On the other hand, 
the burden of proof under \eqref{eq:HT_noncompliant} is borne by the AI provider. 
That is, 
the AI provider is incentivized to give the auditor enough evidence to convince the auditor to reject the null hypothesis $J_0$, 
e.g., the employer is incentivized to prove that their hiring process is not biased.
In this way, 
the choice of hypothesis test should reflect the desired legal presumption and corresponding placement of evidentiary burden. 
Note, however, that this choice may vary across contexts. For example, if the auditor's evidentiary burden is too great under \eqref{eq:HT_compliant}, and the one may wish to shift the evidentiary burden via \eqref{eq:HT_noncompliant}.

\subsection{Hypothesis testing procedure}\label{sec:HT_proc}

In this section, we describe the procedure for casting an AI audit as a hypothesis test. We refer readers interested in hypothesis testing to \citet{casella2002statistical} for a textbook treatment.
For the remainder of this work,
we adopt a presumption of compliance as given in \eqref{eq:HT_compliant},
though our results can be equivalently applied to \eqref{eq:HT_noncompliant}. We discuss four main components of hypothesis testing next: the evidence, decision rule, model, and tolerance. 
\begin{enumerate}
\item\textbf{Evidence.} The auditor has access to evidence $\mathcal{E}$,
as defined in \Cref{sec:setup}.
The auditor is generally limited in the amount of evidence they can gather, for example, due to strictly controlled access to algorithm $f$ or its training data, or due to limited resources.
In the context of NYC Local Law 144, evidence is the historical or test data, including how the automated employment decision tool behaves on such datasets. 

\item\textbf{Decision rule.}
Given evidence $\mathcal{E}$, 
the auditor's goal is to develop an audit (which, in hypothesis testing, is called a     ``decision rule'' $\hat{H}$) that maps evidence $\mathcal{E}$ to a decision $H_0$ or $H_1$.
These two decisions, respectively, correspond to deciding that $f$ is compliant or non-compliant. 
Under \eqref{eq:HT_compliant},
the auditor adopts the default decision $H_0$ unless the evidence is convincing enough for the auditor to reject $H_0$, 
as we review next. 

\item\textbf{Design criteria \& tolerance.}
A decision rule $\hat{H}$ is evaluated based on two quantities: 
the false positive rate (FPR) and true positive rate (TPR), $
    \textsc{FPR} = \mathbb{P}\left(\hat{H} = H_1 | H = H_0\right)$ and $
    \textsc{TPR} = \mathbb{P}\left(\hat{H} = H_1 | H = H_1\right)$,
where $\mathbb{P}$ is taken with respect to randomness in the evidence $\mathcal{E}$ and decision rule $\hat{H}$.
(Observe that the true negative rate and false negative rate can be computed directly from  the FPR and TPR.)
Hypothesis testing is largely concerned with finding rules that maximize the TPR while minimizing the FPR.
Although we do not review them here, 
one approach is to restrict the maximum allowable FPR (known as the \emph{significance level}) to $\zeta$
and find the decision rule that achieves the maximum TPR among all rules with an FPR no more than $\zeta$ and for all possible algorithms in $\cF$.
This rule is known as the uniformly most powerful (UMP) test and can be treated as an ideal benchmark (though it does not always exist). 
The allowable FPR can be viewed as the \emph{tolerance} of an audit.

\hspace{6pt}
Thus, one interpretation of ``statistical significance'' for NYC Local Law 144 is to perform the UMP test given a pre-specified tolerance.
However, there may be more fundamental issues with the casual use of ``statistical significance'' in NYC Local Law 144 for the same reasons discussed at the end of \Cref{sec:burden_proof}.
In particular,
NYC Local Law 144 mandates a ``statistically significant'' audit.
However, ``statistical significance'' generally  applies only to the ability to \emph{reject} the null hypothesis; one cannot obtain a statistically significant \emph{confirmation} of the null hypothesis. 
As such, ``statistical significance'' cannot be achieved if one adopts the null hypothesis  in \eqref{eq:HT_compliant} and the null hypothesis is indeed true. 
Therefore, asking that enough historical or test data be provided until the audit is statistically significant may be problematic, and more careful consideration of the hypothesis test may be needed.
Stated differently, statistical significance is asymmetric, and ambiguity in its intended use (e.g., ambiguity in the choice of null) can result in significantly different interpretations. 

\item\textbf{Model.}
Given evidence $\mathcal{E}$, the auditor's job is to determine whether $g(f) \leq 0$.
Doing so necessarily requires some assumptions (i.e., a \emph{model}) about how $f$ generates $\mathcal{E}$. For example, the auditor may assume that $x$ are drawn from some known distribution $\mathcal{D}$. By definition, the auditor does not know the algorithm $f$ that they wish to audit, but the auditor may assume that $f$ belongs to some model family $\bar{\cF} \subset \cF$. In this case, the model is determined by $(\mathcal{D}, \bar{\cF})$. Without a model, the auditor lacks the necessary assumptions to inform a decision rule; moreover, both \textsc{FPR} and \textsc{TPR} cannot be defined without a model. Crucially, the model clarifies what additional \emph{information}, if any, is needed to conduct an audit. 
In the context of NYC Local Law 144, the key component of the model that the auditor requires is the (empirical) test distribution; 
given this distribution, it is not clear that anything stronger than black-box access is needed since the law seeks to audit for disparate \emph{impact} (i.e., the outcome and not the process) and the metrics of interest are rates (which can be estimated from simple sampling protocols).
In other more complex cases (e.g., privacy or ``unlearning'' audits \cite{cao2015towards,shi2024muse}), additional knowledge of the model and its corresponding assumptions may be needed.

\end{enumerate}
Therefore, to conduct an audit, the auditor takes the following steps: decide on an appropriate model and tolerance, develop a decision rule,  gather evidence, and apply the decision rule.

%% file: sections/challenges.tex
\section{Limitations and future work}\label{sec:limitations}

\emph{Multiple testing.}
While our work considers testing a single property of a model, auditors are typically interested in auditing for multiple criteria or running repeated audits over time. This can present additional challenges, as (1) the reuse of data across audits will invalidate basic statistical guarantees and (2) even audits run on independent samples will not (on their own) control the family-wise error rate or false discovery rate \citep{controlling-fdr-1995}. These issues are exacerbated when the number of audits is not known ex-ante, and may depend on the results of prior audits.

\emph{Explanations, recourse and the limits of auditing.}
While audits that can be cast as hypothesis test can be powerful tools,
they do not necessarily indicate what the provider should do to correct or mitigate these issues. Indeed, as argued in \citet{black-box-insufficient-2024} and discussed in \Cref{sec:black_box}, it is possible that white- and gray-box approaches can be more informative in this regard. For example, hypothesis tests would not necessarily reveal whether the model made a mistake or otherwise behaved unreasonably on a \emph{specific instance}. In such cases, a more localized (or ``counterfactual-based'') approach to auditing might be appropriate \citep{exception-2022, black-box-auditing-2022, counterfactual-metrics-2022}.
Furthermore, black-box auditing does not necessarily enable appropriate recourse when an individual is harmed by an algorithm, as the result of a black-box audit does reveal the reasoning behind a developer's design choices or their intentions.

\emph{Active learning for auditing.}
When given black-box access to an algorithm, 
auditors must choose the inputs that they wish to test, 
i.e., the set $S$ on which to run $f$. The simplest methods involve sampling instances independently from some population of interest, or otherwise specifying $S$ \emph{a priori}. However, a natural approach would also consider choosing these instances in an \emph{online} fashion, in which successive samples are chosen conditional on the output of prior queries. An online sampling procedure will naturally improve the power (i.e., the true positive rate) of an audit at any fixed false positive rate. This class of algorithms is sometimes referred to as \emph{active learning} \cite{hanneke2013statistical}.

\emph{Manipulation-proofness.}
The auditor may also have other concerns, such as ensuring that the audit is manipulation-proof. 
This direction is concerned with removing loopholes that may permit AI developers or companies to pass audits in practice without satisfying the desired criteria in spirit \cite{active-fairness-2022} or guaranteeing that the evidence gathered during an audit reflects how the AI system behaves when deployed \cite{waiwitlikhit2024trustless}.

%% file: sections/tech_related_work.tex
\section{AI auditing techniques}\label{sec:ai_audit_review}

\paragraph{Background: Audit studies.} Audit studies have a long history in the social sciences.
\citet{labor-market-discrimination-2004} find in a study of labor market discrimination that resumes with White-sounding names receive $50$ percent more callbacks, on average, than identical resumes with African-American-sounding names. 
This evidence was gathered by submitting a set of fictitious resumes in response to real help-wanted ads, allowing researchers to experimentally manipulate the perceived race of job applicants in a way akin to the black-box audits described in this work. 

Taking inspiration from this tradition, there is now a growing body of literature that audits algorithmic systems for evidence of consumer harm. Investigators have employed audit studies to examine self-preferencing in search results \citep{google-degrade-search-2016, design-search-engine-2016, google-top-result-2020, google-gatekeeper-2023}, discrimination in online platforms \citep{online-ad-delivery-2013, methods-discrimination-2014, race-effects-ebay-2015, ad-privacy-2015, sharing-econ-2015, cents-on-dollar-2016, mans-wikipedia-2015, bias-freelance-2017, image-representation-2021}, and the effects of algorithmic personalization (particularly on political polarization) \citep{location-personalization-2015, personalization-search-2017, political-search-results-2019, amplification-twitter-2022, like-minded-sources-fb-2023, counterfactual-bots-2023}. A key challenge is that the inputs to these systems are often highly complex, and may not be directly manipulable by researchers. This motivates other causal identification strategies, e.g., by identifying natural experiments in observational data (see \citet{ci-survey-2020} for a recent survey or \citet{mostly-harmless-2008, causality-pearl-2009, ci-imbens-2015} for textbook treatments). For additional background on audit studies, including the legal and ethical questions that arise, as well as recommendations for best practices, we refer to \citet{audit-studies-survey-2021}. For systematic reviews of the algorithm auditing literature, we refer to \citet{lit-review-audits-2021} and \citet{mapping-algorithm-auditing-2024}.

\paragraph{Frameworks for algorithmic auditing.} Perhaps most closely related to this work are general frameworks for ensuring that algorithms satisfy normative and regulatory constraints. As discussed in \Cref{sec: background}, \citet{ai-accountability-gap-2020} propose a framework which guides the development life cycle of an algorithmic decision pipeline. In contrast, we provide an in-depth discussion of black-box audits, 
propose a way to translate between the law and audit procedure,
and describe an open problem related to query complexity. \citet{sociotechnical-audits-2023} take a different perspective, and instead propose the notion of a \emph{socio-technical} audit to directly study the interplay between algorithms and their users. In particular, a socio-technical audit involves experimentally manipulating the \emph{outputs} of an algorithm---for example, via a browser extension which manipulates search results or social media feeds---to study human components of a system (e.g., how user react or modify their behavior) in addition to algorithmic components. \citet{Farley2023} advocates for government-mandated audits of AI systems, and outlines a regulatory framework intended to standardize the practice of AI auditing. \citet{Farley2023} focuses on the policy aspects of AI auditing; in contrast, our work develops a framework for implementing AI audits. \citet{outsider-oversight-2022} and \cite{audits-the-auditors-2023} survey the current practice of auditing and make recommendations for enabling the effective oversight and regulation of algorithms. In the case of \citet{outsider-oversight-2022}, these lessons are drawn primarily from other fields where third-party (or ``outsider") audits have proven effective.

Finally, \citep{regulating-algorithms-2023} propose a framework for regulating algorithms based on model explanations that are tailored to capture specific model characteristics---for example, racial disparities in the model's predictions---rather than to best explain the model's average performance. We further discuss the relationship of auditing and these interpretability techniques in \Cref{sec: additional related}.

\paragraph{Black-box auditing.} Our work focuses on black-box auditing, where the auditor may only \emph{query} the model, rather than e.g., inspecting source code, model architecture or training procedures. This approach is intended to enable third party oversight of algorithms \citep{outsider-oversight-2022}, even in the face of limited cooperation by algorithm providers \citep{audits-the-auditors-2023}. This aligns with the perspective taken in \citet{framework-regulating-social-media-2023}, which propose auditing procedures for algorithms which curate content on social media platforms. It is also the approach taken in \citet{data-minimization-2021}, who develop algorithms for testing compliance with the GDPR's data minimization principle (that an algorithm uses only ``the minimal information that is necessary for performing the task at hand" \citep{data-minimization-2021}). This is also similar to the perspective taken in \citet{black-box-auditing-2022, counterfactual-metrics-2022}, which propose the use of black-box audits to assess \emph{counterfactuals}. For example, such an audit might ask whether, for a given individual, the algorithmic recommendation changes if the individual's race were different. As we discuss in \Cref{sec:limitations}, these localized audits can be useful for individuals seeking recourse for algorithmic harms. 

Finally, contemporaneous work by \citet{black-box-insufficient-2024} argues that a black-box approach is insufficient for rigorous auditing, and highlight the limitations of black-box queries. These include (1) the difficulty of developing a global understanding of how a system behaves, (2) the inability to study system components separately, (3) the possibility that overly simplistic black-box audits can produce misleading results, (4) the limitations of black-box interpretability methods and (5) the inability to suggest \emph{remedies} when models are noncompliant. 
We share the view that broader access (e.g., to model weights, gradients or source code) can enable more in depth auditing of algorithmic systems, 
and we discuss the benefits and limitations of black-box auditing at length in \Cref{sec:black_box}.
Given the strictly controlled access provided to auditors (see our remark above) and concerns such as privacy, 
our discussion of black-box audits is driven by a desire to explore what can be achieved with black-box access,
which can be supplemented with further access to cover its blind spots.

%% file: sections/additional_related_work.tex
\section{Additional related work}
\label{sec: additional related}

\paragraph{Algorithmic fairness.} Our work is inspired by a large literature on algorithmic fairness. This methodological work is itself inspired by well-publicized instances of real-world algorithmic discrimination (e.g, \citet{recidivism-disparate-impact-2016}). Of particular relevance to our work are the many definitions of fairness which have been proposed, including the notion of individual fairness \citep{fairness-through-awareness-2011}, equalized odds \citep{equality-opportunity-2016}, statistical parity or disparate impact \citep{independency-constraints-2009, classifying-without-discriminating-2009, discrimination-free-classification-2010, disparate-impact-2014, censoring-representations-2015, disparate-mistreatment-2016, removing-sensitive-info-2017, nondiscriminatory-predictors-2017} and calibration \citep{calibration-prognostic-scores-2016, compas-risk-scales-2016}, \citep{fp-fn-fa-2016, fairness-calibration-2017, fairness-criminal-justice-2017}. Choosing a particular fairness measure is highly nontrivial, as imposing fairness constraints generally comes at some cost to model accuracy (\citep{cost-of-fairness-2017}). Furthermore, many seemingly natural definitions of fairness turn out to be incompatible with each other (\citep{fairness-tradeoffs-2016, fairness-calibration-2017}). This motivates alternative approaches to fairness which do not directly alter model training procedures \citep{economic-approach-regulation}.

Our work is most closely related to a smaller but growing literature which develops tests for specific kinds of algorithmic harms or failures. For example, \citet{fliptest-2019, active-fairness-2022, inference-fairness-auditing-2023} develop tests for disparities in performance on important (and perhaps legally protected) subgroups, \citet{auditing-individual-fairness-2020} and \citet{inference-individual-fairness-2021} propose algorithms to detect violations of individual fairness, \citet{fairtest-2015} and \citet{auditing-indirect-influence-2016} develop methods to understand how protected attributes influence model behavior (including indirectly). \citet{auditing-expertise-2023, indistinguishability-2024} propose tests to detect whether algorithms fail to incorporate contextual information which may be available to a human decision maker, and \citet{algorithmic-accountability-2019} propose a framework for detecting `input' or proxy discrimination. For additional background we refer to \citet{fairness-frontiers-2018} and \citet{fairness-survey-2019} for surveys of the literature. 

\paragraph{Explainable machine learning.} Our work is also related to a large and growing literature on \emph{explainable} (or interpretable) machine learning. Although we cannot provide a complete overview here, notable works include \emph{LIME} \citep{lime-2016}, a technique for providing explanations for individual model predictions via black-box access, and \emph{SHAP} \citep{shap-2017}, a technique for attributing individual model predictions to specific inputs (`features'). \citet{visualizing-cnn} propose a method for visualizing intermediate layers of a convolutional neural network. These works are broadly motivated by a desire to understand \emph{why} and \emph{how} machine learning models (particularly nonlinear models) make predictions. For additional background, including the challenges of defining model interpretability, we refer to \citet{interpretability-mythos-2016}. For a survey and book-length treatment of specific techniques for model interpretability, we refer to \citet{interpretability-survey, molnar-interpretability-2022}, respectively.

\paragraph{Adversarial attacks.} Finally, our work on black-box auditing is complementary to a rich literature on adversarial machine learning, which seeks to discover (or mitigate against) adversarial inputs---often small perturbations of non-adversarial inputs---which `fool' an algorithm into producing incorrect or incoherent outputs. Indeed, the robustness of algorithmic predictors to adversarial attacks is itself a natural property of interest for both internal and external auditors. Furthermore, the task of \emph{generating} adversarial inputs using a sequence of black-box queries is very similar to the problem of auditing for extreme values, and both are naturally addressed via the machinery of online convex optimization. 

Work on adversarial attacks against machine learning models dates to early email spam filters \citep{adversarial-classification-2004, adversarial-learning-2005, good-word-attacks-2005}. Much of the more recent literature on the vulnerability of deep neural networks to adversarial attacks can be traced to \citet{intriguing-properties-2013}, who document the sensitivity of neural networks to imperceptible perturbations of their inputs. Notable work on adversarial attacks of deep neural networks includes \citet{easily-fooled-2014, explaining-adversarial-examples-2014, adversarial-at-scale-2016, evasion-attacks-2017, decision-based-attacks-2017, black-box-attacks-2018}. To address these vulnerabilities, \citet{adversarial-resistance-2017} propose an approach for training adversarially robust neural networks. More recently, \citet{ignore-prev-prompt-2022, universal-vulernability-2022, black-box-foundation-model-2023} propose techniques for generating adversarial \emph{prompts} for modern foundation models. For additional background on adversarial machine learning, we refer to \citet{wild-patterns-2017}.